\documentclass[preprint]{elsarticle}
\usepackage{epsfig}
\usepackage{amssymb}
\usepackage{amsmath}
\usepackage{amsfonts}
\usepackage{amssymb}
\usepackage{graphicx}
\usepackage{amsthm}
\usepackage{hyperref}
\hypersetup{colorlinks,citecolor=blue,filecolor=blue,linkcolor=blue,urlcolor=blue}
\journal{Physics Letters A}
\begin{document}%
\begin{frontmatter}

\title{Quantum critical points in tunneling junction of topological superconductor
and topological insulator}
\author[Haust,NJU]{Zheng-Wei Zuo\corref{Zuo}}
\ead{zuozw@163.com}
\cortext[Zuo]{Corresponding author.Tel:+8618337991815}
\author[Haust]{Da-wei Kang}
\author[Haust,NJU]{Zhao-Wu Wang}
\author[Haust]{Liben Li}
\address[Haust]{School of Physics and Engineering, Henan University of Science and Technology,
Luoyang 471003, China}
\address[NJU]{National Laboratory of Solid State Microstructures, Nanjing University,
Nanjing 210093, China}

\begin{abstract}
The tunneling junction between one-dimensional topological superconductor and
integer (fractional) topological insulator (TI), realized via point contact,
is investigated theoretically with bosonization technology and renormalization
group methods. For the integer TI case, in a finite range of edge interaction
parameter, there is a non-trivial stable fixed point which corresponds to the
physical picture that the edge of TI breaks up into two sections at the
junction, with one side coupling strongly to the Majorana fermion and
exhibiting perfect Andreev reflection, while the other side decouples,
exhibiting perfect normal reflection at low energies. This fixed point can be
used as a signature of the Majorana fermion and tested by nowadays
experiment techniques. For the fractional TI case, the universal low-energy
transport properties are described by perfect normal reflection, perfect
Andreev reflection, or perfect insulating fixed points dependent on the
filling fraction and edge interaction parameter of fractional TI.

\end{abstract}


\end{frontmatter}

\section{introduction}

Recently, the study of topological superconductors which support Majorana
fermion excitations has been a focus of theoretical and experimental studies
in condensed matter physics\cite{Franz15RMP,Alicea12RPP,Beenakker12ARCMP}.
Majorana fermions being their own anti-particles have exotic non-Abelian
braiding statistics and great potential in the applications of fault-tolerant
topological quantum computation\cite{Nayak08RMP}. There are many proposals
which allow us to engineer topological superconductor (TSC), based on
proximity coupling to $s$-wave superconductors. These include topological
insulators\cite{FuL08PRL,FuL09PRB}, semiconductor quantum
wires\cite{OregY10PRL,LutchynRM10PRL}, and chains of magnetic
adatoms\cite{Nadj-Perge13PRB,PientkaF13PRB,KlinovajaJ13PRL,BrauneckerB13PRL,VazifehMM13PRL}%
. Among these proposals, the most promising candidate for the experimental
realization is the semiconductor quantum wires
proposal\cite{Franz15RMP,StanescuTD13JPCM}. The experimental evidences of
Majorana fermions have been shown in spin-orbit coupled quantum wire
model\cite{Mourik12SCI,DasA12NTP,DengMT12NanoLett}. All other proposals are
being actively pursued\cite{HartS14NTP,Nadj-Perge14SCI}.

Because of these intrinsically fascinating of Majorana fermions, there are
many interesting transport properties and critical points when TSC couples to
other
materials~\cite{LawKT09PRL,ZazunovA11PRB,HutzenR12PRL,FidkowskiL12PRB,Affleck13JSM,KomijaniY14PRB,LutchynRM13PRB,VasseurR14PRX,LeeYW14PRB,ZazunovA14NJP,AltlandA14PRL,ErikssonK14PRL,ChaoSP15PRB,PikulinDI16PRB}%
. A junction between a TSC and a Fermi lead (or interacting lead) is predicted
to exhibit perfect Andreev reflection at low
energies\cite{LawKT09PRL,FidkowskiL12PRB}. Further, a novel type of quantum
frustration and quantum critical points appear at low energies when
one-dimensional (1D) TSC couples to two interacting leads or an interacting
lead with two channels\cite{FidkowskiL12PRB,Affleck13JSM,KomijaniY14PRB}. At
this critical point, the perfect Andreev reflection occurs in one interacting
lead (one channel) and perfect normal reflection in the other. The tunneling
junction between a TSC with chiral Majorana liquid at the edge and a helical
Luttinger liquid is studied\cite{LeeYW14PRB}, the main conclusion of which is
that at low energies, the helical Luttinger liquids is cut into two separated
half wires by backscattering potential and the tunneling between the Majorana
liquid and the helical Luttinger liquid is forbidden. The perfect Andreev
transmission (the reflected hole goes into a different lead from where the
electron arrived) can occur when the edge of topological insulator (TI)
contacts with a Kramers pair of Majorana fermions in
TSC\cite{PikulinDI16PRB}.

Usually, the quantum wires with electron-electron interaction are described by
Luttinger liquids theory\cite{GogolinAO98Book,GiamarchiT04Book} and the
low-energy physics of the tunneling junctions between TSC and interacting
quantum wires are analyzed by renormalization group
method\cite{FidkowskiL12PRB,Affleck13JSM,KomijaniY14PRB,LutchynRM13PRB,VasseurR14PRX,LeeYW14PRB,ZazunovA14NJP,AltlandA14PRL,ErikssonK14PRL}%
. These interacting quantum wires are topological trivial systems. In
contrast, the interplay of the TSC and other topological matters may result in
novel and interesting transport properties. Recently, we have studied the
point contact tunneling junction between 1D TSC and single-channel quantum
Hall (QH) liquids\cite{ZuoZW16EPL}. For the $\nu=1$ integer QH liquid, the
perfect Andreev reflection with quantized zero-bias tunneling conductance
$2e^{2}/h$ is predicted to occur at zero temperature and voltage, which is
caused by Majorana fermion tunneling not by the Cooper-pair tunneling. The
quantized conductance can serve as a definitive fingerprint of a Majorana
fermion. However, for the Laughlin fractional QH liquid cases, the universal
low-energy transport is governed by the perfect normal reflection fixed point
with vanishing zero-bias tunneling conductance.

The edges states of
two-dimensional (2D) integer TI, known as helical liquid, are topologically
protected by time-reversal symmetry. The localized Majorana modes emerge at
interface of superconductor-ferromagnet junction on the edge of 2D
TI\cite{FuL09PRB,BenjaminC10PRB}. The different geometries of TSC coupling
with the edge of 2D TI have been
investigated\cite{LeeYW14PRB,ChaoSP15PRB,PikulinDI16PRB}. The fractional TI
\cite{Bernevig06PRL,Levin09PRL,NeupertT11PRB,NeupertT15PS,MaciejkoJ15NTP,SternA16ARCMP},
which is the strongly interacting version of 2D TI, can be regarded as the
generalization of the fractional QH liquids to time-reversal-invariant
systems. The simplest case of a fractional TI consists of two decoupled copies
of a Laughlin fractional QH states with opposite spin polarizations.
The parafermions (fractionalizing Majorana fermions) can be obtained at the interface
between a SC and a ferromagnet along the edge of fractional TI\cite{David13NTC,LindnerNH12PRX,ChengM12PRB,VaeziA13PRB}.
Due to these intriguing and exotic properties, it is of both theoretical and
practical interest to investigate the transport properties of junction between
the TSC and integer (fractional) TI.

The content of the paper is organized as follows. In Sec. \ref{theory}, using
bosonization technology and renormalization group methods, we firstly research
the tunneling transport signatures of 1D TSC and integer TI. In a finite range
of edge interaction parameter, the edge of TI breaks up into two sections at
the junction, with one side having perfect Andreev reflection due to Majorana
fermion tunneling, while the other side decouples, having perfect normal
reflection. This physical picture of our setup can be tested by present
experimental techniques. Next, we calculate the phase diagram of the
fractional TI case. In Sec. \ref{conclusion}, we make discussions and
concluding remarks.

\section{Theory and Discussion\label{theory}}

\begin{figure}[ptbh]
\centering{ \includegraphics[scale=1.2]{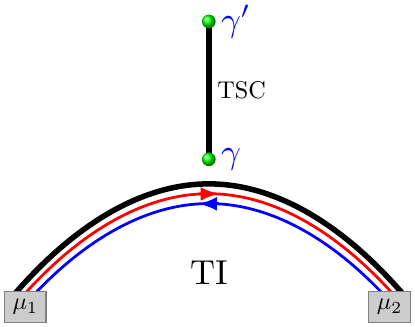}}\caption{
Schematic illustration of the tunneling junction between TSC and TI. The edge
of TI can be described in terms of two bosonic fields $\phi_{\alpha}$. The 1D
TSC is characterized by the Majorana fermions $\gamma$ and $\gamma^{\prime}$.}%
\label{Fig1}%
\end{figure}

In this section, we consider the point contact tunneling junction of 1D TSC
and filling fraction $\nu=1/m$ ($m$ is an odd integer) fractional TI, as shown
in Fig. \ref{Fig1}. When $m=1$, the fractional TI degenerates into 2D
topological insulator. Next, we will use TI to denote the integer and
fractional TI, except where confusion might result from these abbreviations.

The 1D TSC is characterized by the two Majorana fermions $\gamma$ and
$\gamma^{\prime}$ at end points, which can be obtained by a spin-orbit coupled
quantum wire subjected to a magnetic field and proximate to an s-wave
superconductor\cite{OregY10PRL,LutchynRM10PRL}. We assume all the important
energy scales are smaller than the superconducting energy gap and the 1D TSC
is sufficiently long so that Majorana fermion $\gamma^{\prime}$ do not couple
to electrons in the TI. The fractional TI we analyze consists of two coupled
fractional QH states, in which electrons of spin up form a Laughlin fractional
QH states with filling fraction $\nu_{\uparrow}=1/m$ and electrons of spin
down form a Laughlin fractional QH states with filling fraction $\nu
_{\downarrow}=-1/m$. The edge states of the TI are helical Luttinger liquid
and the top edge of TI connects leads $\mu_{1}$ and $\mu_{2}$. Here, we label
the right and left sides of junction by $x>0$ and $x<0$ respectively, and
assume that the two leads are infinitely far away.

The Hamiltonian of the tunneling junction can be expressed as%
\begin{equation}
H=H_{0}+H_{T}%
\end{equation}
where $H_{0}$ is the Hamiltonian of TI edge theory and $H_{T}$ tunneling Hamiltonian.

Firstly, we discuss the edge theories (helical Luttinger liquids) of integer
TI\cite{WuCJ06PRL,XuCK06PRB} and fractional
TI\cite{Levin09PRL,NeupertT11PRB,BeriB12PRL}. Here, we express these theories
within a unified framework. When $m=1$, these reduce to integer TI case. The
edges of TI can be described by two chiral bosonic quantum fields\ $\phi
_{\alpha}$ and the density operators are
\begin{equation}
\rho_{\alpha}=\frac{1}{2\pi}\partial_{x}\phi_{\alpha}%
\end{equation}
where $\alpha=R$ (right-mover with spin up), $L$ (left-mover with spin down).

The boson fields\ $\phi_{\alpha}$ satisfy the Kac-Moody commutation relations%
\begin{equation}
\left[  \phi_{\alpha}\left(  x\right)  ,\phi_{\beta}\left(  x^{\prime}\right)
\right]  =\left(  \sigma_{z}\right)  _{\alpha\beta}\frac{i\pi}{m}%
sgn(x-x^{\prime})
\end{equation}

Because of the time-reversal symmetry, the Hamiltonian of the edge of the TI
is%
\begin{equation}
H_{0}=\int dx\left[  \pi m\upsilon_{F}\left(  \rho_{R}^{2}+\rho_{L}%
^{2}\right)  +2g_{2}\rho_{R}\rho_{L}+g_{4}\left(  \rho_{R}^{2}+\rho_{L}%
^{2}\right)  \right]
\end{equation}
where $g_{2}$ and $g_{4}$ are the amplitudes for dispersion and forward
scattering processes.

To simplify our derivation, we introduce the fields%
\begin{equation}
\varphi=\frac{1}{2}\left(  \phi_{R}+\phi_{L}\right)  ,\theta=\frac{1}%
{2}\left(  \phi_{R}-\phi_{L}\right)
\end{equation}

According to the theory of Luttinger
liquids\cite{GogolinAO98Book,GiamarchiT04Book}, we can express the Hamiltonian
as%
\begin{equation}
H_{0}=\frac{mu}{2\pi}\int dx\left[  K\left(  \partial_{x}\theta\right)
^{2}+\frac{1}{K}\left(  \partial_{x}\varphi\right)  ^{2}\right]
\end{equation}
with%
\begin{align*}
K  &  =\sqrt{\frac{\pi m\upsilon_{F}+g_{4}-g_{2}}{\pi m\upsilon_{F}%
+g_{4}+g_{2}}}\\
u  &  =\sqrt{\left(  1+\frac{g_{4}}{\pi m\upsilon_{F}}\right)  ^{2}-\left(
\frac{g_{2}}{\pi m\upsilon_{F}}\right)  ^{2}}%
\end{align*}
where $K<1$ ($K>1$) for repulsive (attractive) edge interaction, and $K=1$
corresponds to a noninteracting edge. For the noninteracting edge, the
fractional TI can be substituted by a simple electron-hole bilayer where the
two layers are in a Laughlin fractional QH states with filling fraction
$\nu=\pm1/m$.

The electron creation operators can be expressed as%
\begin{equation}
\Psi_{\alpha}^{\dagger}\left(  x\right)  =\Gamma_{\alpha}e^{im\left(
\sigma_{z}\right)  _{\alpha\alpha}\phi_{\alpha}}%
\end{equation}
with $\Gamma_{\alpha}$ the Klein factor that is used to ensure the correct
anti-commutation relations between different fermion species and obey the
following commutation relations.
\begin{equation}
\Gamma_{\alpha}^{\dagger}=\Gamma_{\alpha},\left\{  \Gamma_{\alpha}%
,\Gamma_{\beta}\right\}  =2\delta_{\alpha\beta},\left\{  \Gamma_{\alpha
},\gamma_{\beta}\right\}  =0
\end{equation}

From the first relation above, we can view Klein factors as additional
Majorana fermions, which is important for studying related Majorana fermion
models as shown in \cite{BeriB13PRL,Alexander13PRL}. The third equation ensure
the anti-commutation relations of electrons $\Psi_{\alpha}$ and Majorana
fermion $\gamma$.

As regards the various couplings at the point contact, there are three major
types of tunneling processes: the Majorana fermion-induced tunneling, the
Cooper pairs tunneling, and backscattering of electrons in the helical
Luttinger liquid. The tunneling Hamiltonian of our system at $x=0$ can be
expressed as
\begin{align}
H_{T}=  &  \gamma\sum_{\alpha}t_{\alpha}\left[  \Psi_{\alpha}-\Psi_{\alpha
}^{\dagger}\right]  +\Delta\left[  \Psi_{R}^{\dagger}\Psi_{L}^{\dagger
}+H.c\right] \nonumber\\
&  +u\left[  \Psi_{R}^{\dagger}\Psi_{L}+H.c\right]
\end{align}
where the first term stands for Majorana fermion coupling to electrons of TI,
the second term is the tunneling of Cooper pairs between TSC and TI. The
second term $\Delta$ is the $s$-wave Cooper-pair tunneling induced locally in
the wire by the superconducting pairing\cite{FidkowskiL12PRB,Alicea11NTP},
similar to the tunnel junction between conventional superconductor and
multicomponent fractional QH liquids\cite{KimEA04PRL,ZuoZW14SSC}. The third
term $u$ is the backscattering of electrons in the helical Luttinger liquid.
As supposed in Ref. \cite{LeeYW14PRB}, the backscattering of electrons can
occur because of the broken time-reversal symmetry by the TSC. For
time-reversal invariant TSC, the electrons backscattering is
forbidden\cite{PikulinDI16PRB}.

After bosonization, the tunneling term becomes%
\begin{align}
H_{T}=  &  2i\gamma\sum_{\alpha}t_{\alpha}\Gamma_{\alpha}\cos\left[  m\left(
\varphi\left(  0\right)  \pm\theta\left(  0\right)  \right)  \right]
+2\Delta\sin\left[  2m\theta\left(  0\right)  \right] \nonumber\\
&  +2u\sin\left[  2m\varphi\left(  0\right)  \right]
\end{align}
where the upper (lower) sign is for $\alpha=R$ $(L)$. The first tunneling term
factorize into Klein-Majorana interaction and charge sector parts. We can
define ordinary fermion $\psi_{\alpha}=\left(  \gamma+i\Gamma_{\alpha}\right)
/2$ with $\left\{  \psi_{\alpha},\psi_{\alpha}^{\dagger}\right\}  =1$, so%
\begin{equation}
i\gamma\Gamma_{\alpha}=2\psi_{\alpha}^{\dagger}\psi_{\alpha}-1=\pm1
\end{equation}
and we can see that the values correspond to this energy level being occupied
and empty. Consequently, the Klein-Majorana fusion procedure can eliminate the
Majorana degrees of freedom\cite{BeriB13PRL,Alexander13PRL} and simplify our
theoretical calculations.

We first assume $t_{\alpha}$, $\Delta$ and $u$ are weak perturbation and study
these fate using perturbative renormalization group (RG) analysis. First, we
pass to a Lagrangian formalism by a Legendre transform of the Hamiltonian and
integrate out the bosonic fields in the partition function of the system
except at $x=0$, then obtain a theory defined only at the location of the
point contact. Next, the bosonic field $\varphi$ ($\theta$) are split into
slow ($s$) and fast ($f$) modes: $\varphi_{s}\left(  \tau\right)
=\int_{-\Lambda/b}^{\Lambda/b}\frac{d\omega}{2\pi}e^{-i\omega\tau}%
\varphi(\omega)$ and $\varphi_{f}\left(  \tau\right)  =\int_{\Lambda
/b<\left\vert \omega\right\vert <\Lambda}\frac{d\omega}{2\pi}e^{-i\omega\tau
}\varphi(\omega)$, with $\Lambda$ as an energy cutoff, $b>1$ as a scale
factor, and $\tau=it$ the Euclidean time. The $\theta_{s}$ and $\theta_{f}%
$\ have a similar definition. Third, we integrate over the fast modes and the
new effective action for the slow degrees of freedom can be calculated using
cumulant expansion to the lowest-order approximation. Last, the effective
action is identical in its structure to the original action hence is
renormalizable and the next step is rescaling. In a word, the lowest-order
flows for the coupling $t_{\alpha}$, $\Delta$, and $u$ are%
\begin{equation}
\frac{dt_{\alpha}}{d\ln b}=t_{\alpha}\left(  1-\frac{m}{4}\left(  K+\frac
{1}{K}\right)  \right)  \label{tFTI-RG}%
\end{equation}%
\begin{equation}
\frac{d\Delta}{d\ln b}=\Delta\left(  1-\frac{m}{K}\right)
\label{CopperFTI-RG}%
\end{equation}%
\begin{equation}
\frac{du}{d\ln b}=u\left(  1-mK\right)  \label{backscattering-RG}%
\end{equation}

In what follows, we first analyze the phase diagrams of integer TI coupling to
1D TSC and calculate its transport signatures. Then, the fractional TI case is investigated.

\subsection{Topological insulator case}

When $m=1$, our system reduces to the 2D integer TI (quantum spin Hall effect)
case. From Eq.(\ref{tFTI-RG}), we can see that
when $2-\sqrt{3}<K<2+\sqrt{3}$, the Majorana fermion coupling to electrons
term becomes relevant. For $K>1$, the Cooper-pair tunneling term becomes
relevant (from Eq.\ref{CopperFTI-RG}). When $K<1$, the electrons backscattering is relevant (from Eq.\ref{backscattering-RG}). So, when
$1<K<2+\sqrt{3}$, the Majorana fermion coupling and Cooper-pair tunneling are
in competition with each other. For $2-\sqrt{3}<K<1$, the Majorana fermion
coupling and electrons backscattering are in competition with each other.
Thus, the tunneling junction will exhibit fascinating phase diagrams.
According to these RG flows of Eqs. (\ref{tFTI-RG}-\ref{backscattering-RG}),
for $K>\sqrt{3}$, the Cooper-pair tunneling is the dominant scattering process
and the low-energy transport is controlled by perfect Andreev reflection fixed
point with quantized zero-bias conductance $2e^{2}/h$. For $K<1/\sqrt{3}$, the
electrons backscattering is most relevant and\ the system flows to the perfect
insulating fixed point at low energies, analogous to the result of Ref.
\cite{LeeYW14PRB}. This fixed point corresponds to the physics picture
where the backscattering cuts the edge of TI into two halves (denoted by right $R$
and left $L$ sides) and the conductance between the leads $\mu_{1}$ and
$\mu_{2}$ is zero. The perfect insulating fixed point in our setup and the
fixed point $I$ in Ref. \cite{LeeYW14PRB} are the same, because the edge in each setup is effectively cut
into two separated pieces by electron backscattering at the quantum point
contact. There are no currents flowing through the edge of TI and we have two half-infinite helical Luttinger liquids.
Next, we first analyze the stability of the perfect insulating fixed point for
$K<1/\sqrt{3}$, and then research the scattering properties for $1/\sqrt
{3}<K<\sqrt{3}$, where the Majorana fermions coupling is most relevant.

\subsubsection{Phase diagram}

For convenience, we regard the left and right edge parts of 2D TI at $x=0$ as
the leads $L_{\mu}$ and $R_{\mu}$. To investigate the stability of the perfect
insulating fixed point for $K<1/\sqrt{3}$, we consider the possible
perturbations around it. Now, our system is similar to 1D TSC coupling to two
interacting leads or one lead with two
channels\cite{FidkowskiL12PRB,Affleck13JSM,LeeYW14PRB}, while the differences
from our system are that the two leads are the edges of 2D TI and have the
same Luttinger interaction parameter $K$. So, we can use these part conclusion
for our setup. Physically, the deviations from this fixed point mean that the
single electron transmission between left and right sides at $x=0$ is allowed.
There are other possible perturbation around it such as Majorana fermion
tunneling $t_{\beta}$ ($\beta=R_{\mu}, L_{\mu}$) and Cooper-pair tunneling $\Delta_{\beta}$. The
single electron transmission $\lambda$ can be expressed as
\begin{equation}
H_{\lambda}=\lambda\left(  \Psi^{\dagger}\left(  0^{+}\right)  \Psi\left(
0^{-}\right)  +H.c\right)
\end{equation}

\begin{figure}[ptbh]
\centering{ \includegraphics[scale=1.2]{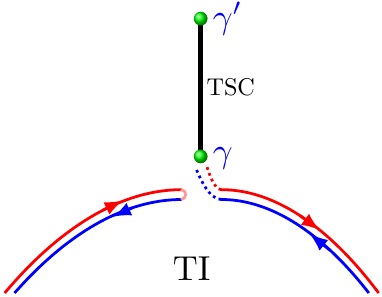}}\caption{
Schematic illustration picture of $A\otimes N$ or
$N\otimes A$ fixed points.}%
\label{Fig2}%
\end{figure}

Based on the above RG analysis, we can derive the scaling dimensions of
$t_{\beta}$, $\Delta_{\beta}$, and $\lambda$, yielding $D\left(  t_{\beta
}\right)  =1/\left(  2K\right)  $, $D\left(  \Delta_{\beta}\right)  =2/K$, and
$D\left(  \lambda\right)  =1/K$. So, the Cooper-pair tunneling and single
electron transmission are irrelevant. However, the Majorana fermion tunneling
is relevant for $1/2<K<1/\sqrt{3}$ and destabilize the perfect insulating
fixed point. We may naively conclude that the Majorana fermion is
simultaneously and strongly coupling with the two sides under renormalization
and the low-energy transport is controlled by perfect Andreev reflection fixed
points in the two sides (label $A\otimes A$). However, this $A\otimes A$ fixed
point is unstable, as shown in Ref \cite{Affleck13JSM}. First, the
$A\otimes A$ quantum critical point occurs only when $t_{1}$ and $t_{2}$
exactly balance. But, the Majorana fermion has a spin
structure\cite{SticletD12PRL,SedlmayrN15PRB}. Generally, Majorana fermion
$\gamma$ couples with one side stronger than the other. Second, using the
`$\epsilon$-expansion' techniques\cite{Affleck13JSM}, we can derive that the
bigger tunneling $t_{\beta}$ grows under renormalization and the smaller flow
to zero. So, the system flows to $A\otimes N$ or $N\otimes A$ fixed points
which corresponding to perfect Andreev reflection with quantized zero-bias
tunneling conductance $2e^{2}/h$ in one side and perfect normal reflection
with vanishing zero-bias tunneling conductance in the other. Here, we discuss
the stability of the $A\otimes N$ or $N\otimes A$ fixed points. For
simplicity, we assume the system flows to $A\otimes N$ fixed point where
perfect Andreev reflection occurs in left side and perfect normal reflection
takes place in the right side. The possible perturbations around this fixed
point are that the electrons backscattering $\Psi_{R}^{\dagger}\left(
0^{-}\right)  \Psi_{L}\left(  0^{-}\right)  +H.c$ in left side with scaling
dimension $2K$, Cooper-pair tunneling $\Psi_{R}^{\dagger}\left(  0^{+}\right)
\Psi_{L}^{\dagger}\left(  0^{+}\right)  +H.c$ in right side with scaling
dimension $2/K$, and electrons transmission $\Psi^{\dagger}\left(
0^{+}\right)  \Psi\left(  0^{-}\right)  +H.c$ between the two sides with
scaling dimension $(K+1/K)/2$. All these perturbations are irrelevant. Thus,
for $1/2<K<1/\sqrt{3}$, the low-energy physics is described by $A\otimes N$ or
$N\otimes A$ fixed points where the edge of TI breaks up into two sections at
the junction, with one side coupling strongly to the Majorana mode and
exhibiting perfect Andreev reflection, while the other side decouples,
exhibiting perfect normal reflection. The illustrative picture of $A\otimes N$ or
$N\otimes A$ fixed points is shown in Fig.\ref{Fig2}. In addition, for $K<1/2$, the perfect
insulating fixed point is stable.

Next, we analyze the scattering processes for the region of $1/\sqrt
{3}<K<\sqrt{3}$. Now, the Majorana fermions coupling is the leading relevant
operator. Under renormalization, the Majorana fermion couples more strongly
with the two right and left movers (chiral bosonic fields $\phi_{\alpha}$).
This might indicate the system flows toward $A\otimes A$ fixed point. As
discussed above, this $A\otimes A$ fixed point is also unstable. The reason is
that the critical point $A\otimes A$ needs PT (parity $\otimes$ time reversal)
symmetry protecting, while the electrons backscattering $\Psi_{R}^{\dagger
}\Psi_{L}+H.c.$ term breaks time reversal symmetry, as shown in the Appendix
$F$ in Ref \cite{Affleck13JSM}. By way of the `$\epsilon$-expansion'
techniques, we found that the system flows to $A\otimes N$ or $N\otimes A$
fixed points which corresponding to perfect Andreev reflection in one mover
and perfect normal reflection in the other. According to the perfect Andreev
reflection boundary condition $\Psi_{R}^{\dagger}\left(  0^{\pm}\right)
=-\Psi_{L}\left(  0^{\pm}\right)  $, the edge of TI breaks up into two
sections at the junction, with one side coupling strongly to the Majorana mode
and exhibiting perfect Andreev reflection, while the other side decouples,
exhibiting perfect normal reflection. This result is the same as the case for
the region of $1/2<K<1/\sqrt{3}$. So, the $A\otimes N$ or $N\otimes A$ fixed
points are stable for $1/\sqrt{3}<K<\sqrt{3}$. In summary, we can derive the
phase diagram of the integer TI case, as shown in table \ref{table1}. For $K>\sqrt{3}$, the
low-energy physics is governed by the perfect Andreev reflection fixed point
with Cooper-pair tunneling. For $1/2<K<\sqrt{3}$, there exist $A\otimes N$ or $N\otimes A$
fixed points where the edge of TI breaks up into two sections at
the junction, perfect Andreev reflection with quantized zero-bias
tunneling conductance $2e^{2}/h$ occurs in one side and perfect normal reflection
with vanishing zero-bias tunneling conductance occurs in the other. For
$K<1/2$, the low-energy physics is determined by perfect insulating fixed point with electrons backscattering.

\begin{table}[ptbh]
\caption{Phase diagram for the point tunneling junction of 1D TSC and integer
TI}%
\label{table1}
\centering
\begin{tabular}{ll}%
\hline
Region & Fixed point\\
\hline
$K>\sqrt{3}$ & perfect Andreev reflection\\
$1/2<K<\sqrt{3}$ & $A\otimes N$ or $N\otimes A$\\
$K<1/2$ & perfect insulating\\
\hline
\end{tabular}
\end{table}

\subsubsection{Differential conductance}

In what follows, we calculate the low energy physical observable conductance
of our setup for $K<1$ with physical interest. For convenience, the TSC, leads
$L_{\mu}$ and $R_{\mu}$ (the left and right sides for the edge of TI) are
labeled as $0$, $1$, and $2$. For the $K<1/2$, the leading irrelevant
scattering processes are electron transmission and Majorana fermion tunneling.
The tunneling conductance $G$ between TSC and TI, and conductance $G_{12}$
between leads $L_{\mu}$ and $R_{\mu}$ to the lowest-order approximation in
infinitesimal voltage $V$ or temperature $T$ are%
\begin{align}
G\left(  V\right)   &  \sim V^{1/K-2},G\left(  T\right)  \sim T^{1/K-2},\\
G_{12}(V)  &  \sim V^{2/K-2},G_{12}\left(  T\right)  \sim T^{2/K-2}.
\end{align}

For the $1/2<K<1$, we assume first that the system flows toward $A\otimes N$
fixed point and then calculate the differential conductance. The results for
$N\otimes A$ fixed point are straightforward. Now, the perfect Andreev
reflection occurs at the left side and perfect normal reflection occurs at the
right side. The conductance $G_{01}$ between TSC and lead $L_{\mu}$,
conductance $G_{02}$ between TSC and lead $R_{\mu}$, and conductance $G_{12}$
between leads $L_{\mu}$ and $R_{\mu}$ in bias voltage $V$ are%
\begin{align}
G_{01}\left(  V\right)   &  \sim\left(  \frac{2e^{2}}{h}-c_{V}V^{4K-2}\right)
,\label{G01}\\
G_{02}\left(  V\right)   &  \sim V^{4/K-2},\\
G_{12}(V)  &  \sim V^{K+1/K-2}, \label{G12}%
\end{align}
where $c_{V}$ is a non-universal constant. The low temperature dependences of
these conductance have similar power-law scaling behaviors.

When the 1D TSC is in topological trivial phase and Majorana fermion tunneling
is absent, the tunneling conductance between TSC and 2D TI, and conductance
$G_{12}$ between leads $L_{\mu}$ and $R_{\mu}$ for $K<1$ are%
\begin{align}
G\left(  V\right)   &  \sim V^{4/K-2},G\left(  T\right)  \sim T^{4/K-2}%
\label{GtivialTC}\\
G_{12}\left(  V\right)   &  \sim V^{2/K-2},G_{12}\left(  T\right)  \sim
T^{2/K-2} \label{Gtivial12}%
\end{align}
where the Cooper-pair tunneling and electrons transmission terms are the
leading irrelevant operators.

Finally, we discuss the experimental measurements for our setup. The
experimental evidences of TSC have been shown in InSb quantum
wire\cite{Mourik12SCI,DasA12NTP,DengMT12NanoLett}. On the other hand, the 2D
TI has been observed in HgTe/CdTe\cite{Konig07SCI} and
InAs/GaSb\cite{KnezI11PRL} quantum wells. Here, we take the HgTe/CdTe for
example. It is rather nontrivial to estimate the experimentally edge
interaction parameter $K$ of 2D TI. As calculated in Refs.
\cite{HouCY09PRL} and \cite{StromA10PRL}, the value of interaction
parameter for HgTe/CdTe quantum well is $0.5<K<0.55$. So, in the case of
HgTe/CdTe quantum well, the low-energy physics is determined by the non-trivial
stable $A\otimes N$ or $N\otimes A$ fixed points. The different bias voltage
dependences in Eqs. (\ref{G01}-\ref{Gtivial12}) can identify whether the TSC
is nontrivial or not. So, the quantized conductance and different power-law
scaling behaviors in Eqs. (\ref{G01}-\ref{G12}) can be used as a
signature for the Majorana fermion in 1D TSC. For the InAs/GaSb
quantum well, the Fermi velocity of the edge modes can be controlled by gates
and Luttinger interacting parameter can be fine tuned\cite{LiTX15PRL}, we can use the
InAs/GaSb quantum well to test the phase diagram of integer TI coupling to 1D
TSC as a function of the Luttinger interacting parameter.

\subsection{Fractional topological insulator case}

From the Eqs. (\ref{tFTI-RG}-\ref{backscattering-RG}), we can see that for all
odd integer $m>1$, the tunneling term $t$ of Majorana fermion coupling to
electrons is irrelevant. When $K>m$, the Cooper-pair tunneling term $\Delta$
is relevant and the low-energy transport is controlled by perfect Andreev
reflection fixed point with quantized zero-bias conductance $2e^{2}/h$. When
$1/m<K<m$, the low-energy physics is governed by the perfect normal reflection
fixed point with a vanishing zero-bias tunneling conductance. For $K<1/m$, the
electrons backscattering is relevant and the system flows to the perfect
insulating fixed point at low energies. This fixed point corresponds to the
physics picture where the backscattering cuts the edge into two halves and the
conductance between the leads $\mu_{1}$ and $\mu_{2}$ is zero.

Now we analyze the stability of the perfect insulating fixed point. There are
three types of scattering processes that might destabilize the fixed point:
Majorana fermion tunneling, Cooper-pair tunneling, and the quasiparticles
$\psi_{R/L,\alpha}\left(  x\right)  =\digamma_{\alpha}e^{i\phi_{\alpha}\left(
x\right)  }$($\digamma_{\alpha}$ is the Klein factor) transmission. The same
RG analysis also applies to this perfect insulating fixed point. The scaling
dimensions of quasiparticles transmission, Majorana fermion coupling, and
Cooper-pair tunneling are given by $1/(mK)$, $m/(2K),$ and $2m/K$. We can see
that for $K<1/m$, these three processes are all irrelevant and this fixed
point is stable for strong repulsive interaction. Next, we calculate the
corrections to the tunneling conductance $G$ and two-terminal conductance
$G_{12}$ between leads $L_{\mu}$ and $R_{\mu}$ due to finite bias voltage or
temperature. The scaling behaviors of the tunneling conductance $G$ and
two-terminal conductance $G_{12}$ to the lowest-order approximation are%
\begin{align}
G\left(  V\right)   &  \sim V^{m/K-2},G\left(  T\right)  \sim T^{m/K-2}\\
G_{12}(V)  &  \sim V^{2/mK-2},G_{12}\left(  T\right)  \sim T^{2/mK-2}%
\end{align}

In the perfect normal reflection regime, when $\sqrt{3}<K<m$, the leading
irrelevant operator around this point is Cooper-pair tunneling term $\Delta$.
The tunneling conductance $G$ at bias voltage $V$ (low temperature $T$) has
the power-law form
\begin{equation}
G\left(  V\right)  \sim V^{2m/K-2},G\left(  T\right)  \sim T^{2m/K-2}%
\end{equation}
and when $1/m<K<\sqrt{3}$, the leading irrelevant operator is Majorana fermion
tunneling term and the tunneling conductance $G$ is given by
\begin{equation}
G\left(  V\right)  \sim V^{m\left(  K+1/K\right)  /2-2},G\left(  T\right)
\sim T^{m\left(  K+1/K\right)  /2-2}%
\end{equation}

The Majorana fermion-induced tunneling is irrelevant which can be rooted in
strong repulsive interaction of fractional TI. The quasiparticles of
fractional TI have fractional charge and obey fractional exchange statistics
which are forbidden for tunneling. In a word, the universal low-energy
transport properties is described by perfect Andreev reflection fixed point
for $K>m$, perfect normal reflection fixed point for $1/m<K<m$, perfect
insulating fixed point for $K<1/m$. When the 1D TSC is trivial, we can derive
the same phase diagram. But, the tunneling conductance has different power-law
behaviors when the bias voltage or temperature is nonzero. The power-law form
of tunneling conductance $G$ for perfect insulating regime ($K<1/m$) is
\begin{equation}
G\left(  V\right)  \sim V^{4m/K-2},G\left(  T\right)  \sim T^{4m/K-2}%
\end{equation}

In the perfect normal reflection regime ($1/m<K<m$), the tunneling conductance
$G$ is%
\begin{equation}
G\left(  V\right)  \sim V^{2m/K-2},G\left(  T\right)  \sim T^{2m/K-2}%
\end{equation}

\section{Conclusion\label{conclusion}}

In this paper, we have studied the point contact tunneling junction between 1D
TSC and integer (fractional) TI with bosonization and renormalization group
methods. For the junction of 1D TSC and topological insulator, according to
the strength of edge interaction parameter, there is a non-trivial stable
fixed point that the edge is cut into two halves, the perfect normal
reflection occurs in one side, and perfect Andreev reflection occurs in the
other side. We can use the InSb quantum wire and HgTe/CdTe quantum well to
test this non-trivial stable fixed point, which can be used to identify the
Majorana fermion in 1D TSC. The $A\otimes N$ ($N\otimes A$) and perfect
insulating fixed points appear in the setups of Refs. \cite{Affleck13JSM}
and \cite{LeeYW14PRB} respectively, however, the edge states of the TI are
topologically protected. For the fractional TI case, the universal
low-energy transport properties are described by perfect normal reflection,
perfect insulating, or perfect Andreev reflection fixed points dependent on
the filling fraction and edge interaction parameter of fractional TI. From the
above results, we found that there are different fixed points for the
different topological matters coupling to TSC. When other strongly correlated
topological matters couple with TSC, the exotic electron tunneling transport
properties and critical points maybe occur.

\textbf{Acknowledgment} This work was supported by the National Natural
Science Foundation of China under grant numbers 11447008 (Z.Z.W.), 11204066
(D.W.K.) 11404095 (Z.W.W.).


\end{document}